\documentclass{sf2a-conf2024}
\usepackage{graphicx}
\usepackage{hyperref}
\usepackage[]{natbib}  
\usepackage{epstopdf}
\usepackage{amsmath}
\usepackage{sidecap}

\def\BibTeX{{\rm B\kern-.05em{\sc i\kern-.025em b}\kern-.08em
    T\kern-.1667em\lower.7ex\hbox{E}\kern-.125emX}}
\bibpunct{(}{)}{;}{a}{}{,}  
\pdfoutput=1
\defcitealias{Tokuno2022AsteroseismologyOscillations}{TT22}

\begin{document}

\TitreGlobal{SF2A 2024}


\title{Constraining core-to-envelope differential rotation in gamma-Doradus stars from inertial dips properties}

\runningtitle{Constraining differential rotation in gamma-Doradus stars from inertial dips properties}

\author{L. Barrault}\address{Institute of Science and Technology Austria (ISTA), Am Campus 1, 3400 Klosterneuburg, Austria}

\author{S. Mathis}\address{Université Paris-Saclay, Université Paris Cité, CEA, CNRS, AIM, 91191, Gif-sur-Yvette, France}

\author{L. Bugnet$^{1}$}




\setcounter{page}{237}


\maketitle


\begin{abstract}
   The presence of dips in the gravito-inertial modes period-spacing pattern of $\gamma$-Dor stars is now well established by recent asteroseismic studies. Such Lorentzian-shaped inertial dips arise from the interaction of gravito-inertial modes propagating in the radiative envelope of intermediate-mass main sequence stars with pure inertial modes that propagate in their convective core. 
   We aim to investigate the signature of a differential rotation between the convective core and the near-core region inside $\gamma$-Dor stars from the inertial dip properties.
   We first describe the bi-layer rotation profile we use and the approximations we adopt to maintain the analyticity of our study.
   We then describe our results on the inertial dip formation, location, and shape. We derive a modified Lorentzian profile and we compare it to the previously obtained results in the solid-body rotation case.
  This work highlights the inertial dips' probing power of the convective core rotation, an important observable in the context of the understanding of the angular momentum transport and chemicals mixing inside stars.
\end{abstract}



\section{Introduction}
\par During the Main Sequence (MS), gravito-inertial (g-i) modes are unique probes of the stellar radiative zone. The key relevant observable is the period-spacing pattern (PSP): the evolution of the spacing between modes of consecutive radial orders but the same horizontal structure with the period of the mode.
The slope of the PSP in an inertial frame has been used to measure rotation rates at the basis of the radiative envelope, a location at which g-i modes reach the highest sensitivity \citep{VanReeth2015DetectingStudies,Ouazzani2017AStars,Christophe2018DecipheringStars}. 
This complements the measurement of surface rotation rates to provide two probing points at the limits of the radiative envelope. 
A key remaining question is how to probe the properties of the convective core, as they are important for the internal mixing and magnetism of the star. In this framework, $\gamma$-Dor stars are ideal targets for such an analysis, with 611 stars in \textsl{Kepler} data showing prograde g-i modes \citep{Li2020Gravity-modeKepler}.
\par \citet{Ouazzani2020FirstRevealed} and \citet{Saio2021RotationModes} proved that pure inertial modes propagating in the convective core of the star couple through the convective-radiative boundary with g-i modes. This results in a characteristic dip in the PSP, later observed in 16 $\gamma$-Dor stars \citep{Saio2021RotationModes}. The location of the dip in period, as well as its width and depth, were proven to depend on stellar parameters and evolution.
To provide a detailed physical understanding of the formation of the dip, \cite{Tokuno2022AsteroseismologyOscillations} (hereafter \citetalias{Tokuno2022AsteroseismologyOscillations}) described theoretically the coupling mechanism and found a Lorentzian shape for the dip.
\citet{Galoy2024PropertiesStars} further conducted a thorough numerical study of the inertial dip formation,
deriving empirical relations for the width and the location of the dips as a function of the density stratification of the convective core and the near-core stratification gradient.

\par 
While \citet{Ouazzani2020FirstRevealed},\citetalias{Tokuno2022AsteroseismologyOscillations} and \citet{Galoy2024PropertiesStars} remained in the framework of solid-body rotation, \citet{Saio2021RotationModes} allowed a steep differential rotation profile from the convective core to the near-core regions.
\citet{Moyano2024AngularStars} further used these results to put constraints on the angular momentum redistribution processes at play in the radiative interior of $\gamma$-Dor stars.
For that matter, it is key to finely explore the detectability of differential rotation from the dip properties. We describe in this work an extension of the \citetalias{Tokuno2022AsteroseismologyOscillations} formalism and give the analytical derivation of the dip shape in the case of differential rotation.
We conclude by giving hints on the impact of differential rotation on the dip detectability, and foresee future complementary work pursued in the companion study Barrault et al. (under review).
  
\section{Summary of the analytical study}
\begin{figure}[ht!]
 \centering
 \includegraphics[width=0.35\textwidth,clip]{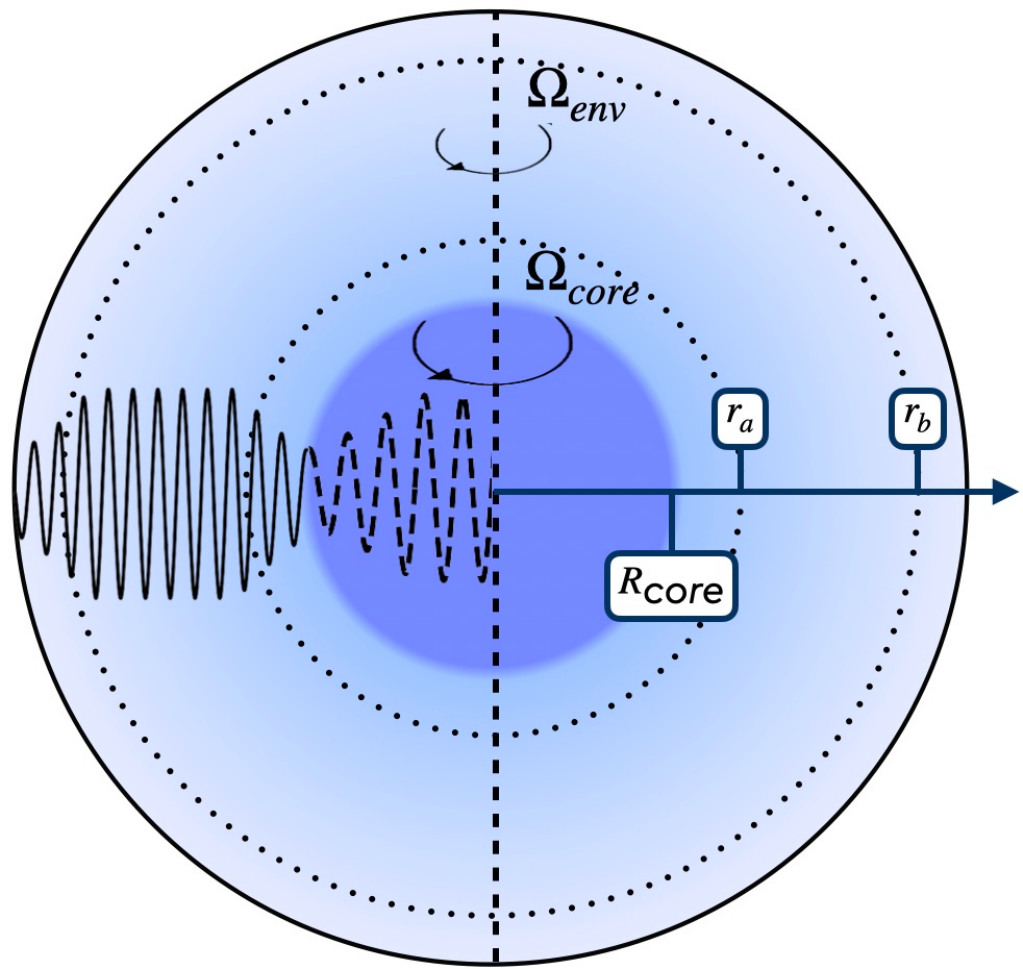}    
  \caption[\textwidth]{Sketch of the chosen bi-layer rotation profile: $\Omega_{\mathrm{core}}$ for $r<R_{\mathrm{core}}$, $\Omega_{\mathrm{env}}$ for $r>R_{\mathrm{core}}$. g-i modes are propagating in the zone $r\in[r_{\mathrm{a}};r_{\mathrm{b}}]$, being evanescent in the region $r\in[R_{\mathrm{core}};r_{\mathrm{a}}]$. Pure-inertial modes are trapped in the convective core $r<R_{\mathrm{core}}$. The matching of the relevant quantities is done at the location $r=R_{\mathrm{core}}$.}
  \label{Barrault:fig1}
\end{figure}

\subsection{Outline of the model}
To provide a theoretical understanding of the effect of differential rotation on the inertial dip properties, we adopt a bi-layer rotation model, with a convective core rotating at $\Omega_{\mathrm{core}}$ and a radiative envelope at $\Omega_{\mathrm{env}}$, such as $\Omega_{\mathrm{env}} = \alpha_{\mathrm{rot}} \Omega_{\mathrm{core}}$. We consider the interaction at the boundary between gravito-inertial (g-i) modes propagating in the envelope and pure inertial modes propagating in the core in the sub-inertial regime. The frequencies of the modes in the frame co-rotating with the respective zones are $\sigma_{\mathrm{env}} = \sigma_{\mathrm{in}} + m\Omega_{\rm env}$ and $\sigma_{\mathrm{core}} = \sigma_{\mathrm{in}} + m\Omega_{\rm core}$, with $\sigma_{\rm in}$ the frequency in an inertial frame. They verify $\sigma_{\mathrm{env}} < 2 \Omega_{\mathrm{env}}$ and $\sigma_{\mathrm{core}} < 2 \Omega_{\mathrm{core}}$. A sketch of the star and its two zones is shown in Fig.~\ref{Barrault:fig1}.
\par We adopt the Traditional Approximation of Rotation (TAR) within the envelope \citep[e.g.][]{Gerkema2008GeophysicalApproximation}. The TAR consists of neglecting the latitudinal component of the rotation vector in the Coriolis acceleration. It is known to hold in highly stably stratified layers of stellar interiors. We thus enforce the condition $N \gg 2\Omega_{\mathrm{core}}$, $N$ being the Brunt-Väisälä frequency. Under these requirements, the TAR solution approaches the solution obtained with a full treatment of the Coriolis acceleration \citep{Prat2016AsymptoticDynamics}. Using as well the Cowling approximation \citep{Cowling1941TheStars}, neglecting the wave's gravific potential fluctuation, the set of equations governing g-i modes becomes separable and the horizontal structure follows Hough functions \citep[e.g.][]{Lee1997Low-frequencyDependence}. G-i modes are denoted by continuous lines in Fig.~\ref{Barrault:fig1}. They propagate between a lower and upper turning point, respectively $r_{a}$ and $r_{b}$. Between the boundary, at the location $R_{\mathrm{core}}$, and $r_{a}$, g-i modes are evanescent. We thus hypothesize a strong gradient of $N$ in the near-core region, maintaining this zone thin enough for g-i modes not to be significantly damped, or diverging from Hough solutions. This hypothesis is reasonable throughout the MS of the star as highlighted by realistic stellar models \citep[e.g.][]{Dhouib2022DetectingField}.
\par In the convective core, the mixing triggered by convection ensures a negligible Brunt-Väisälä frequency, discarding the use of the TAR. The equations governing the pure inertial mode structure are separable in the case of solid-body rotating full spheres of constant density \citep{Wu2005OriginModes}. We thus choose as a first step a uniform averaged density profile in the core, for the study to remain analytical. Under this framework, pure inertial modes can decompose in Bryan solutions \citep{Bryan1889TheEllipticity}. They are denoted by dashed oscillations in Fig.~\ref{Barrault:fig1}.
\par The structure of modes propagating from both sides of the boundary depends on the spin parameters related to each of the zones, $s_{\rm core} = {2\Omega_{\rm core}}/{\sigma_{\rm core}} $ and $s_{\rm env} = {2\Omega_{\rm env}}/{\sigma_{\rm env}}$. This is a key difference between our study and the solid-body rotating case as the equality of core and envelope mode frequencies implied the equality of spin parameters in the latter case. The spin parameter of a pure inertial mode in an isolated convective core $s^{*}_{\rm core}$ is set by an eigenvalue problem originating from a fixed boundary condition of null radial displacement ($\xi_{r} = 0$) at the core boundary. For the $(l=3, m=-1)$ pure inertial mode, producing the dips observed in the PSP of $(k=0, m=-1)$ prograde g-i modes \citep{Saio2021RotationModes}, this conditions sets the spin parameter of a pure inertial mode at $s_{core}^{*} = 11.32$.

\subsection{Analytical formulation and relevant parameters}
In contrast to the solid-body rotating case considering a single co-rotating frame, we need to ensure the equality of the frequencies in an inertial frame. This translates into a relation between the spin parameters in the two zones, $s_{\mathrm{env}} = G(s_{\mathrm{core}})$, with the function $G$ defined as:
\begin{equation}
    G(s) = \frac{\alpha_{\mathrm{rot}}s}{1+\frac{m}{2}(\alpha_{\mathrm{rot}}-1)s} \, .
\end{equation}
The reasoning for the matching of the modes at $R_{\rm core}$ is similar to the one held in \citetalias{Tokuno2022AsteroseismologyOscillations}. We ensure the continuity of the radial displacement and the Lagrangian pressure perturbation from both sides of the boundary:
$\xi_{r,\mathrm{core}}|_{R_{\rm core}} = \xi_{r,\mathrm{env}}|_{R_{\rm core}}$ and $\delta p_{\mathrm{core}}|_{R_{\rm core}} = \delta p_{\mathrm{env}}|_{R_{\rm core}}$.
The assumed near-core strong gradient of $N$ allows for an expansion of the solutions from $r_{\mathrm{a}}$ to $R_{\mathrm{core}}$, using a (small) coupling parameter $\epsilon$ decreasing with increasing near-core $N$ gradient. The matrices $\mathcal{M}$ and $\mathcal{N}$ introduced in equations (51) and (52) of \citetalias{Tokuno2022AsteroseismologyOscillations} now depend on two spin parameters: $s_{\mathrm{core}}$ and $s_{\mathrm{env}}$. Considering the most prominent mode interaction allows us to simplify the (infinite) matrix problem into a single coupling equation. This reads, sticking to \citetalias{Tokuno2022AsteroseismologyOscillations} notations:

\begin{equation}
    \frac{C_{l}^{m}(1/s_{\text{core}})}{P_{l}^{m}(1/s_{\text{core}})}\times \frac{\sigma_{\text{env}}^{2}}{\sigma_{\text{core}}^{2}} \times \frac{Y_{k}(s_{\text{env}})}{X_{k}(s_{\text{env}})} \simeq \epsilon \, ,
    \label{eq:coupling}
\end{equation}
where we highlight on the left-most fraction the contribution of the core pure inertial mode, and on the right-most one the contribution of envelope g-i modes.
With further manipulations, we find that the effect of differential rotation on the dip structure can be caught into a modified Lorentzian structure, described in Table \ref{tab:formula}.

\begin{table}[ht]
    \centering
    \begin{tabular}{c|c}
         \citet{Tokuno2022AsteroseismologyOscillations} & This work \\
        \hline
          $\frac{1}{\Delta P} - \frac{1}{\Pi_0} \simeq \frac{\Gamma/\pi}{(P-P_{*} + \Gamma/\sqrt{3})^{2}+\Gamma^{2}}$ &     $\frac{1}{\Delta P} - \frac{1}{\Pi_0} \simeq \frac{\Gamma_{\text{diff}}/\pi\frac{\mathrm{d}G^{-1}}{\mathrm{d}s}\big|_{\bar{s}}}{\left((P-P_{*})\frac{\mathrm{d}G^{-1}}{\mathrm{d}s}\big|_{\frac{\bar{s}+s^{*}}{2}} + \frac{\Gamma_{\text{diff}}}{\sqrt{3}}\right)^{2}+\Gamma_{\text{diff}}^{2}}$ 
    \end{tabular}
    \caption{Analytical formulation of the dip profile ({\bf Right}) and comparison to the solid-body rotating case ({\bf Left}). $\Delta P$ is the period spacing between two consecutive modes, $\Pi_0$ the buoyancy travel time, $\Gamma_{\mathrm{diff}} = \frac{3\pi\epsilon}{4\Omega_{\mathrm{env}}V_{\mathrm{diff}}}$ reducing to $\Gamma$ in the solid-body rotating case, with $V_{\mathrm{diff}}$ a structural factor. $\bar{s}$ is the mean spin parameter between two modes, and $s^{*} = G(s_{\mathrm{core}}^{*})$, the envelope spin parameter corresponding to the core spin parameter of the pure inertial mode and $P_{*} = \pi s^{*}/\Omega_{\mathrm{env}}$. Starred quantities refer to the pure inertial mode only.}
    \label{tab:formula}
\end{table}
\subsection{Effects of differential rotation on the morphology of the inertial dip}


\begin{figure}[ht!]
 \centering
 \includegraphics[width=\textwidth,clip]{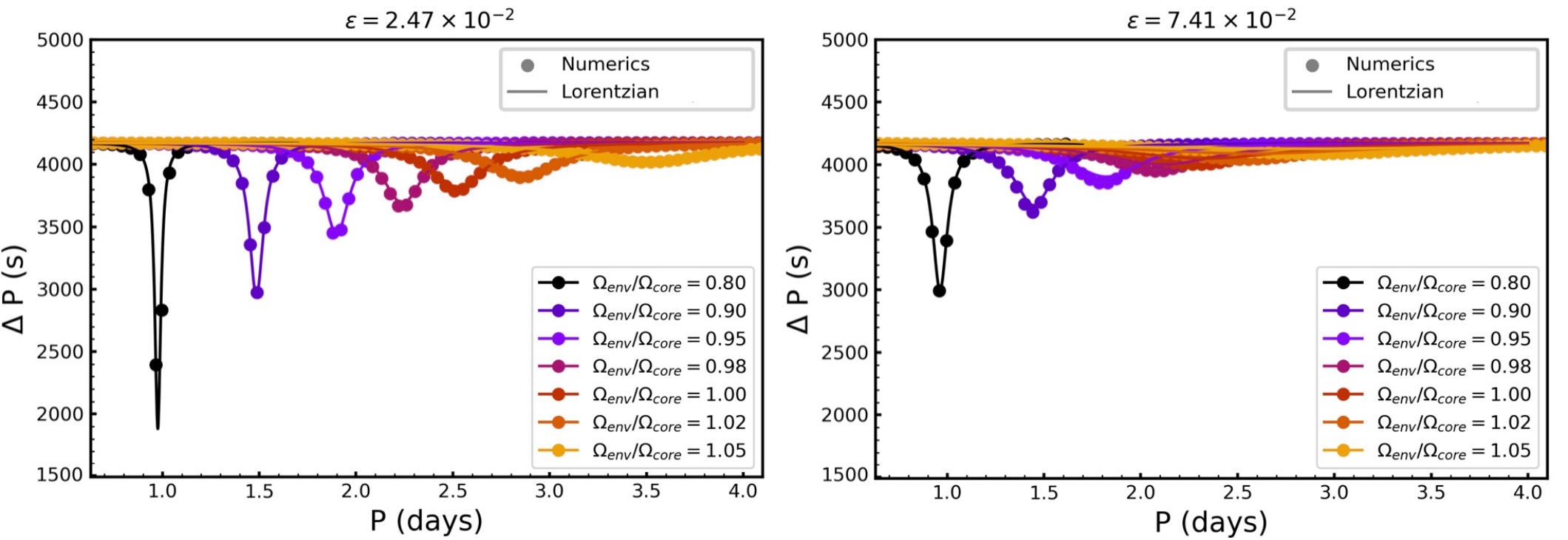}    
  \caption{Dips formed by the coupling between $(k = 0, m = -1)$ g-i modes in the envelope and $(l = 3, m = -1)$ pure inertial modes in the core, in the frame co-rotating with the envelope. For each panel corresponding to low ({\bf Left}) and high ({\bf Right}) values of coupling parameter, PSPs are superimposed on one another, taking values of $\alpha_{\rm rot} = \Omega_{\mathrm{env}}/\Omega_{\mathrm{core}}$ in $[0.80, 1.05]$. We consider $\Omega_{\mathrm{env}}/2\pi = 2.16 \, \mathrm{d}^{-1}$ and $\Pi_{0} = 4175 \, \mathrm{s}$, corresponding to KIC12066947. The analytical result given by the rightmost formula of Table~\ref{tab:formula} is shown in solid lines, with numerical results solving Eq.~\ref{eq:coupling} as circles. }
  \label{Barrault:fig2}
\end{figure}

Inertial dips arise at a period varying with core-to-envelope differential rotation, as shown in Fig.~\ref{Barrault:fig2}. Increasing core rotation compared to the envelope one corresponds to a low period of the dip, while a slower-rotating core increases the dip's period when compared to the case of uniform rotation. At fixed coupling parameter value, with increasing core rotation rate compared to the envelope, the dip gets also thinner and deeper. This is to be compared to the effect of a decrease of the gradient of stratification near-core, hence an increase of $\epsilon$ (Right panel of Fig.~\ref{Barrault:fig2}): inertial dips for the same $\alpha_{\mathrm{rot}}$ become wider and shallower.

\par An increasing core rotation compared to the envelope thus has a similar impact as a decrease of the coupling parameter on the dip shape. For such a regime, Doppler-shifted g-i modes interacting with the pure inertial mode at $s_{core}^{*} = 11.32$ have a lower envelope spin parameter. The spectrum of g-i modes in the frame co-rotating with the core is less dense than in the solid-body rotating case. Thus the number of g-i modes close in period to the pure inertial one is reduced. The coupling is less strong, hence resulting in a thinner dip.
\par The results are thus coherent with the qualitative picture, comprising as well information on the variation of the spatial structure of modes with differential rotation through the parameter $\Gamma_{\mathrm{diff}}$ (see Table~\ref{tab:formula}). They highlight the potential difficulty of detecting the dip structure in a regime of core rotating slower than the envelope: in the right panel, dips in this regime are barely distinguishable from the straight baseline of the PSP. On the contrary, a situation in which the core rotates faster than the envelope would potentially make the dip structure appear clearer in the PSP, this being mitigated by the lower number of modes in the dip. 

\section{Conclusions}
\par We described the effect of core-to-envelope differential rotation on the inertial dip location and shape in the period-spacing pattern of $\gamma$-Dor stars. With a bi-layer rotation profile, an analytical formula can be derived, that gathers the effect of the Doppler shift of g-i modes and the variation of their angular structure. An increasing core rotation decreases coupling strength between core pure inertial modes and g-i modes, thereby increasing the thinness of the inertial dip. A potential impact on the dip detectability is briefly discussed.
\par This framework is a laboratory for the study of the effect of core and near-core internal processes on the inertial dip structure. It highlights the great potential of using this feature as a probe of the innermost regions of the star and specifically the convective core, a location otherwise uneasy to access. It remains to study potential degeneracies between near-core density stratification and differential rotation and to analyze the detectability of core rotation from data. This will be the scope of Barrault et al. (under review).

\begin{acknowledgements}
S.M. acknowledges support from the European Research Council (ERC) under the Horizon Europe programme (Synergy Grant agreement 101071505: 4D-STAR), from the CNES SOHO-GOLF and PLATO grants at CEA-DAp, and from PNPS (CNRS/INSU). While partially funded by the European Union, views and opinions expressed are however those of the author only and do not necessarily reflect those of the European Union or the European Research Council. Neither the European Union nor the granting authority can be held responsible for them.
\end{acknowledgements}

\bibliographystyle{aa}  
\bibliography{Barrault_S01} 

\end{document}